\documentclass{llncs}
\usepackage{graphicx}

\begin{document}

\title{
  AceWiki: Collaborative Ontology Management in Controlled Natural Language
}

\author{
  Tobias Kuhn
}

\institute{
  Department of Informatics, University of Zurich, Switzerland\\
  \texttt{tkuhn@ifi.uzh.ch}\\
  \texttt{http://www.ifi.uzh.ch/cl/tkuhn}
}

\maketitle

\begin{abstract}
AceWiki is a prototype that shows how a semantic wiki using controlled natural language --- Attempto Controlled English (ACE) in our case --- can make ontology management easy for everybody. Sentences in ACE can automatically be translated into first-order logic, OWL, or SWRL. AceWiki integrates the OWL reasoner Pellet and ensures that the ontology is always consistent. Previous results have shown that people with no background in logic are able to add formal knowledge to AceWiki without being instructed or trained in advance.
\end{abstract}

\section{Introduction}

Since ontologies are often defined within communities, semantic wikis could be used for their collaborative creation and management. Unfortunately, most of the existing semantic wikis do not support expressive ontology languages in a general way. They do not allow the users to add complex axioms like ``every landlocked country borders no sea''. Furthermore, the existing semantic wikis are often hard to understand for people who are not familiar with the technical terms of logic and ontologies.

AceWiki\footnote{See \cite{kuhn08acewiki} and \texttt{http://attempto.ifi.uzh.ch/acewiki}} tries to solve both problems by using controlled natural language. Ordinary people who have no background in logic should be able to understand, modify, and extend the formal content of a wiki.

\begin{figure}[tb]
  \begin{center}
    \includegraphics[width=11.5cm]{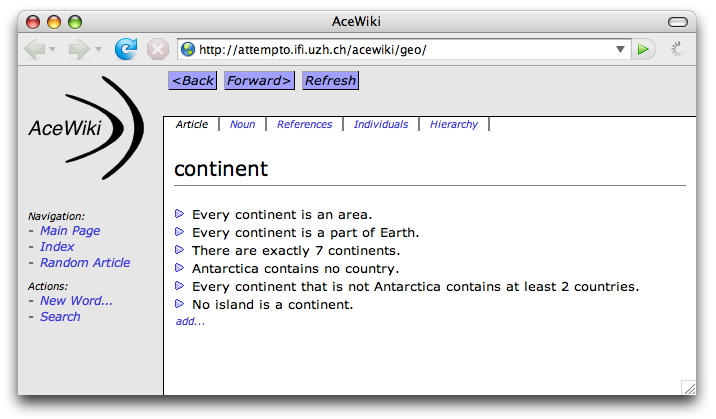}
    \caption{The web interface of the AceWiki prototype}
    \label{fig:window}
  \end{center}
\end{figure}

Many existing semantic wikis are classical wikis enriched with semantic annotations. The goal is not to manage stand-alone ontologies, but rather to give some kind of formal backbone to the wiki articles. We follow a different approach --- similar e.g. to the \textit{myOntology} project \cite{siorpaes07myontology} --- by providing a wiki that is dedicated to building and maintaining ontologies. In contrast to myOntology, we do not restrict ourselves to lightweight (i.e. relatively inexpressive) ontologies. The use of controlled natural language allows us to express also complex axioms in a natural way. Figure \ref{fig:window} shows a screenshot of the AceWiki interface.

In our usage scenario, a community of domain experts uses AceWiki to create and maintain a formal knowledge base in a collaborative manner. There are two exemplary wiki instances --- one about geography and the other about protein interactions --- that demonstrate how AceWiki could be used to represent knowledge of such communities.

AceWiki has been introduced in \cite{kuhn08acewiki}. Since then, several new features have been added, for example the integration of a reasoner and the support for number restrictions (``at most 3'', ``exactly 5'', etc.).

\section{Attempto Controlled English}

Attempto Controlled English (ACE)\footnote{See \cite{fuchs:flairs2006} and \texttt{http://attempto.ifi.uzh.ch}} is the controlled natural language that is used for AceWiki. ACE appears completely natural since it is a subset of English. Restrictions of the syntax and the definition of a small set of interpretation rules make it a formal language that is automatically translatable into first-order logic. ACE supports a wide range of natural language constructs: singular and plural noun phrases, active and passive voice, relative phrases, anaphoric references, existential and universal quantifiers, negation, modality, and more. In the past, ACE has successfully been applied for different tasks in different research areas, for example as a query language for ontologies \cite{bernstein04talking}, as a knowledge representation language for the biomedical domain \cite{kuhn2006protein}, and as a rule language for a multi-semantics rule engine \cite{kuhn07acerules}.

Furthermore, ACE has been used as a natural language front-end to OWL with a bidirectional mapping of ACE to OWL \cite{kaljurand07phd}. This mapping covers all of OWL 2 except data properties and some very complex class descriptions. AceWiki relies on this work for translating ACE sentences into OWL, which allows us then to do reasoning with existing OWL reasoners.

\section{Design and Evaluation}

The goal of AceWiki is to show that semantic wikis can be more natural and at the same time more expressive than existing semantic wikis.

\begin{figure}[tb]
  \begin{center}
    \includegraphics[width=12cm]{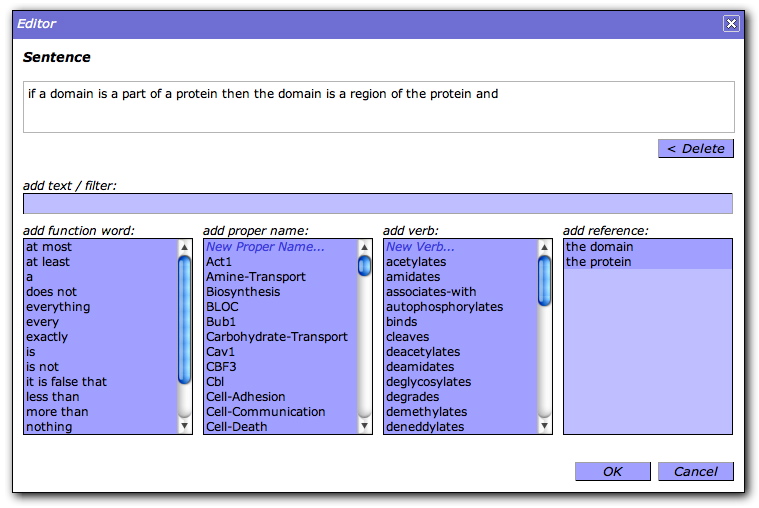}
    \caption{The predictive editor of AceWiki}
    \label{fig:editor}
  \end{center}
\end{figure}

Naturalness is achieved by representing the formal statements in ACE. Since ACE is a subset of natural English, every English speaker can immediately read and understand the content of the wiki. In order to enable easy creation of ACE sentences, AceWiki provides a predictive editor that shows step-by-step the words that are syntactically possible at a given position in the sentence. Figure \ref{fig:editor} shows a screenshot of the predictive editor of AceWiki. Furthermore, the AceWiki interface does not use technical terms like ``ontological element'', ``property'', or ``subclass'' but uses instead terms like ``word'', ``transitive verb'', or ``hierarchy'' which should be much more familiar to people with no background in logic.

AceWiki makes use of the high expressivity of ACE that goes beyond OWL and SWRL. We do not like the idea of cutting down the expressivity just for the sake of reasoning performance. Even if some statements become so complex that it is almost impossible to do reasoning with them, it is better to have them formalized than just left out. We do not lose anything, since we are free to ignore those complex statements for certain reasoning tasks.

In our previous work \cite{kuhn08acewiki}, we conducted a user experiment that proved that ordinary people with no background in logic are able to deal with AceWiki. The participants --- without being instructed how to interact with the interface --- were asked to add knowledge to AceWiki. About 80\% of the created sentences were correct and sensible. This is remarkable since most of the sentences were quite complex: more than 60\% of them contained an implication or a negation or both. Using the predictive editor which the participants had never seen before, they needed on average only five minutes to create their first correct sentence.

\section{Reasoning in AceWiki}

We have started to integrate the OWL reasoner Pellet\footnote{\texttt{http://pellet.owldl.com/}} into AceWiki. Since ACE sentences can be beyond the expressivity of OWL, the reasoner cannot consider all sentences. In order to make this clear to the users, each sentence is tagged as blue (inside of OWL) or red (outside of OWL):
\begin{center}
  \includegraphics[width=9cm]{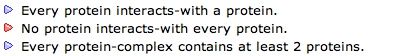}
\end{center}
In this way, it is easy to explain to the users that only the blue statements are considered when the reasoner is used. We plan to provide an interface that allows skilled users to export the formal content of the wiki and to use it within an external reasoner or rule-engine. Thus, even though the red statements cannot be interpreted by the built-in reasoner they can still be useful.

Consistency checking plays a crucial role because any other reasoning task requires a consistent ontology in order to return useful results. Most other semantic wikis do not have this problem since their languages are simply not expressive enough to ever run into inconsistency.

In order to ensure that the ontology is always consistent, AceWiki checks every new sentence --- immediately after its creation --- whether it is consistent with the current ontology. Otherwise, the sentence is not included in the ontology:
\begin{center}
  \includegraphics[width=9cm]{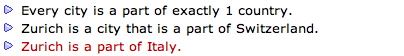}
\end{center}
After the user created the last sentence of this example, AceWiki detected that it contradicts the current ontology. The sentence is included in the wiki article but the red font indicates that it is not included in the ontology. The user can remove this sentence again, or keep it and try to reassert it later when the rest of the ontology has changed.

For this approach, it is very important to perform incremental reasoning which Pellet supports only partially at the moment. For that reason, AceWiki does not scale very well. We expect that future reasoners will be able to run much faster in such incremental scenarios.

Not only asserted but also inferred knowledge can be represented in ACE. At the moment, AceWiki can show inferred class hierarchies and class memberships. Furthermore, we are working on a query feature for AceWiki. Questions will be formulated in ACE and evaluated by the reasoner:
\begin{center}
  \includegraphics[width=9cm]{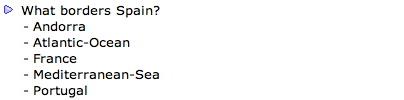}
\end{center}
Thus, ACE can be used not only as an ontology- and rule-language, but also as a query-language.

\section{Conclusions}

The AceWiki prototype shows how ontologies can be managed in a natural way within a wiki. It demonstrates how semantic wikis using controlled natural language can be expressive and easy to use at the same time. Our previous evaluation showed that AceWiki is indeed easy to learn. We explained how AceWiki ensures --- in a very simple way --- the consistency of the ontology which is the basis for other integrated reasoning services.


\begin{thebibliography}{10}

\bibitem{bernstein04talking}
  Abraham Bernstein, Esther Kaufmann, Norbert E. Fuchs, June von Bonin.
  Talking to the Semantic Web --- A Controlled English Query Interface for Ontologies.
  \textit{Proc. 14th Workshop on Information Technology and Systems}, 2004

\bibitem{fuchs:flairs2006}
  Norbert E. Fuchs, Kaarel Kaljurand, Gerold Schneider.
  Attempto Controlled English Meets the Challenges of Knowledge Representation, Reasoning, Interoperability and User Interfaces.
  \textit{Proc. 19th International FLAIRS Conference (FLAIRS'2006)}, 2006

\bibitem{kaljurand07phd}
  Kaarel Kaljurand.
  \textit{Attempto Controlled English as a Semantic Web Language}.
  PhD thesis, Faculty of Mathematics and Computer Science, University of Tartu, 2007

\bibitem{kuhn2006protein}
  Tobias Kuhn, Lo\"{i}c Royer, Norbert E. Fuchs, Michael Schroeder.
  Improving Text Mining with Controlled Natural Language: A Case Study for Protein Interactions.
  \textit{Proc. Third International Workshop on Data Integration in the Life Sciences 2006 (DILS'06)}, Springer, 2006

\bibitem{kuhn07acerules}
  Tobias Kuhn.
  AceRules: Executing Rules in Controlled Natural Language.
  \textit{Proc. First International Conference on Web Reasoning and Rule Systems (RR2007)}, Springer, 2007

\bibitem{kuhn08acewiki}
  Tobias Kuhn.
  AceWiki: A Natural and Expressive Semantic Wiki.
  \textit{Proc. of Semantic Web User Interaction at CHI 2008: Exploring HCI Challenges}, CEUR Workshop Proceedings, 2008

\bibitem{siorpaes07myontology}
  Katharina Siorpaes, Martin Hepp.
  myOntology: The Marriage of Ontology Engineering and Collective Intelligence.
  \textit{Proc. Bridging the Gap between Semantic Web and Web 2.0}, 2007

\end{thebibliography}
\end{document}